\documentclass[twocolumn,preprintnumbers,amsmath,amssymb]{revtex4}
\input{epsf}
\usepackage{graphicx}

\usepackage{dcolumn}
\usepackage{bm}
\usepackage{color}

\definecolor{Black}{named}{Black}
\definecolor{Blue}{named}{Blue}
\definecolor{Red}{named}{Red}

\def\nab{{\hbox{\boldmath$\nabla$}}}
\def\lsim{\,{}_\sim^{<}\,}
\def\gsim{\,{}_\sim^{>}\,}

\begin{document}

\title{The Evolution of the Large-Scale Tail of Primordial Magnetic Fields}

\author{Karsten Jedamzik}
\affiliation{Laboratoire de Physique Theorique et Astroparticules, UMR5207-CNRS, Universite Montpellier II, F-34095 Montpellier, France}

\author{G\"unter Sigl}
\affiliation{{II}. Institut f\"ur Theoretische Physik, Universit\"at Hamburg, Luruper
Chaussee 149, D-22761 Hamburg, Germany}

\begin{abstract}
Cosmic magnetic fields may be generated during early cosmic phase transition,
such as the QCD- or electroweak- transitions. 
The magnitude of the remainder of such fields at the present epoch 
crucially depends on the exponent $n$ of their 
(initially super-Hubble) 
large-scale tail, i.e. $B_{\lambda}\sim \lambda^{-n}$. 
It has been claimed that causality requires $n=5/2$,
contrary to much earlier claims of $n=3/2$. Here we 
analyze this question in detail.
First, we note that contrary to current belief, the large-scale 
magnetic field tail is not established at the phase transition itself, 
but rather continuoulsy evolves up to the present epoch. Neglecting turbulent flows we find
$n=7/2$, i.e. very strongly suppressed large-scale fields. However, 
in the inevitable presence of turbulent flows we find that the large-scale
magnetic field tail has sufficient time to evolve 
to that of the fluid turbulence. 
For white noise fluid turbulence this yields $n=3/2$ up to a certain scale 
and $n=5/2$ beyond for the magnetic field spectrum. This picture is also not
changed when primordial viscosity and fluid flow dissipation is taken into
account. Appreciable primordial magnetic fields originating
from cosmic phase transitions seem thus possible.
\end{abstract}


\maketitle

\section{Introduction}
The origin of galactic- and cluster- magnetic fields is still unknown
(cf.~\cite{Brandenburg:2004jv}). For long, one viable possibility was the generation of magnetic
fields during some early magnetogenesis epoch~\cite{Grasso:2000wj} which than could act
as seed fields for a galactic dynamo~\cite{Kulsrud:1999bg}, or even under optimistic
conditions provide the complete magnetic field after gravitational
collapse~\cite{Banerjee:2003xk}.
Another viable option is the generation of magnetic fields in stars
and the expulsion of these fields into the intergalactic medium after the end
of the life time of stars~\cite{vallee}. In either scenario many open questions
remain. Irrespective of the origin of galactic magnetic fields it is 
interesting to know if substantial primordial relic magnetic fields 
could have survived the evolution of the early Universe.

This is particularly so due to recent claims of a lower limit on the
magnitude of magnetic field strength in the intergalactic medium
of $B\gsim 10^{-16}-10^{-14}$Gauss (depening on coherence length scale)
~\cite{Neronov:1900zz}. These claims are due to  
$\gamma$-ray observations of distant blazars by Fermi and HESS and
detailed analysis of $\gamma$-ray showers on the extragalactic-
and cosmic microwave- background radiation in the presence and absence of
intergalactic magnetic fields (cf.~\cite{Aharonian:1993vz}). 
They seem to indicate that a
large volume fraction $f\gsim 0.5$ of voids a filled by magnetic 
fields which seems only difficult to achieve in scenarios of outflows/winds from
galaxies.~\cite{Dolag:2010ni}

It has been claimed~\cite{Caprini:2001nb}, 
that phase transition generated primordial
magnetic~\cite{Sigl:1996dm} 
fields are too weak to be at the origin of the observed galactic- and cluster-
magnetic fields, even under very optimistic assumptions on galactic dynamo
amplification. The only exception would be here the case of completely
helical fields produced during the QCD transition, nevertheless, requiring
substantial dynamo amplification as well~\cite{Caprini:2009pr}.
The main point here is that due to "causality" arguments, it is claimed,
the spectral index $n$ in
\begin{equation}
\label{eq:n}
\epsilon_B(k)\simeq \frac{1}{8\pi}\int_0^k {\rm d^3}k^{\prime}
{\bf B}^*_{\bf k} {\bf B}_{\bf k} = \epsilon_B^0\,
\biggl(\frac{k}{k_0}\biggr)^{2n}\, ,
\end{equation}
would be large $n\gsim 5/2$ yielding a very blue spectrum and not much
power on scales much above the phase transition Hubble radius.
Here ${\bf B}_{\bf k}$ is the Fourrier transform of magnetic field, $k$ is
wave vector, and $\epsilon_B$ is magnetic energy density.

In this paper we analyze the question of the spectral index in detail. In the
next subsection we compute the magnetic field tail arising from the
superposition of uncorrelated magnetic dipoles in vaccum, exemplifying that
$n=3/2$~\cite{Hogan} is possible even for supper-Hubble distances. In Section
III we replace the vacuum by the highly conductive medium in the early
Universe, finding a suppressed $n=7/2$. In Section IV we include the
realistic existence of fluid flows in our analysis analyzing the
evolution of the large-scale tail first, in the turbulent fluid, and then
in a fluid described by alternating periods of turbulent and viscous 
magnetohydrodynamics (MHD), as shown to occur in 
Ref~\cite{Banerjee:2004df}. In both cases we find
$n = 3/2$ up to some large scale. Conclusions are drawn in Section V.

\section{The Magnetic Field Spectrum due to Uncorrelated Dipoles}
For a magnetic dipole density ${\bf M}$, in the magneto-static
  approximation the magnetic field can be written as
$${\bf B}({\bf r})\propto\int\,d^3{\bf r}^\prime
\frac{3({\bf r}-{\bf r}^\prime){\bf M}({\bf r}^\prime)\cdot({\bf
    r}-{\bf r}^\prime)-{\bf M}({\bf r}^\prime)|{\bf r}-{\bf
    r}^\prime|^2}{|{\bf r}-{\bf r}^\prime|^5}\,,$$
where $\mu_0$ is the magnetic susceptibility in vacuum. From this one
can easily show that
\begin{equation}
\label{eq:dipole}
\left\langle{\bf B}_{\bf k}{\bf B}_{-{\bf k}}\right\rangle\propto
\int\,d^3{\bf r}^\prime \left\langle{\bf M}({\bf r}){\bf M}({\bf
    r}^\prime)\right\rangle\,e^{i{\bf k}\cdot({\bf r}-{\bf r}^\prime)}\,.
\end{equation}
For uncorrelated dipoles, $\left\langle{\bf M}({\bf r}){\bf M}({\bf
    r}^\prime)\right\rangle\propto\delta({\bf r}-{\bf r}^\prime)$,
this gives $\left\langle{\bf B}_{\bf k}{\bf B}_{-{\bf
      k}}\right\rangle=$const and therefore, comparing with Eq.~(\ref{eq:n}), a
slope $\epsilon_B(k)\propto k^3$, thus $n=3/2$. The magneto-static
approximation is good at scales much smaller than the light-travel
distance and thus certainly at Mpc scales, which are much smaller than
the Hubble radius today.

\section{Magnetic Field Evolution neglecting Fluid Flows}

We assume that the total electrical current density consists of an
external source ${\bf j}_{\rm ex}$ and an Ohmic current induced by
the electro-magnetic field,
\begin{equation}\label{eq:current}
  {\bf j}={\bf j}_{\rm ex}+\sigma\left({\bf E}+{\bf v}\times{\bf B}\right)\,,
\end{equation}
with $\sigma$ the conductivity. Here the non-vanishing external current 
${\bf j}_{\rm ex}$ is assumed to be dynamically forced during the
epoch of magnetogenesis.
Upon neglecting the displacement current $\partial_t{\bf E}$
for the total current one has Amp\`eres law $4\pi{\bf j}=\nab\times{\bf B}$.
Writing the external current as the rotation of the dipol-density
{\bf M},
\begin{equation}\label{eq:M}
  {\bf j}_{\rm ex}=\nab\times{\bf M}\,,
\end{equation}
inserting ${\bf E}$ from Eq.~(\ref{eq:current}) into
Faraday's induction equation 
\begin{equation}\label{eq:Faraday}
  \partial_t{\bf B}=-\nab\times{\bf E}
\end{equation}
and using $\nab\cdot{\bf B}=0$ one has
\begin{equation}\label{eq:B}
  \partial_t{\bf B}=\nab\times({\bf v}\times{\bf B})+\eta\left[\Delta{\bf B}-
  4\pi\left(\Delta{\bf M}-\nab\left(\nab\cdot{\bf M}\right)\right)\right]\,,
\end{equation}
where we have used the resistivity $\eta=1/(4\pi\sigma)$.

We now express these equations in the expanding Universe in terms
of redshift $z$ and at the same time perform a Fourier transformation
of the spatial dependence. Using $\partial_z=-1/[(1+z)H(z)]\partial_t$
with $H=\dot a/a$ the Hubble constant and the scale factor $a=1/(1+z)$
and neglecting the fluid velocity term in Eq.~(\ref{eq:B})
one then obtains for the evolution of the Fourier component 
${\bf B}_{\bf k}(z)$
for comoving wave number ${\bf k}$,
\begin{eqnarray}\label{eq:M_k}
  \partial_z{\bf B}_{\bf k}(z)&=&2\,\frac{{\bf B}_{\bf k}(z)}{1+z}+\\
  &&+\eta(z)\frac{1+z}{H(z)}{\bf k}^2\left[{\bf B}_{\bf k}(z)
  -4\pi{\bf M}^\perp_{\bf k}(z)\right]\,,\nonumber
\end{eqnarray}
where ${\bf M}^\perp_{\bf k}(z)\equiv{\bf M}_{\bf k}(z)-
({\bf \hat{k}}\cdot{\bf M}_{\bf k}(z))\, {\bf \hat{k}}$ is the component of
${\bf M}_{\bf k}(z)$ perpendicular to ${\bf k}$, with
$\bf \hat{k}$ the unit vector in direction of $\bf k$. 
The solution of Eq.~(\ref{eq:M_k}) which vanishes
before the source term is switched on is given by
\begin{widetext}
\begin{equation}\label{eq:M_k_solution}
  {\bf B}({\bf k},z)=4\pi{\bf k}^2(1+z)^2\int_z^\infty\exp\left[
  -{\bf k}^2\int_z^{z^\prime}\frac{\eta(\tilde z)(1+\tilde z)}{H(\tilde z)}
  d\tilde z\right]\frac{\eta(z^\prime){\bf M}_\perp({\bf k},z^\prime)}
  {(1+z^\prime)H(z^\prime)}dz^\prime\,.
\end{equation}
\end{widetext}
It is convenient to define a critical wave number as
\begin{equation}\label{eq:k_crit}
  k_r(z)=\left[\int_z^\infty\frac{\eta(\tilde z)(1+\tilde z)}
  {H(\tilde z)}d\tilde z\right]^{-1/2}\,.
\end{equation}
For $k\gg k_r$ the exponential in Eq.~(\ref{eq:M_k_solution})
cuts off contributions from $z+\Delta z\gtrsim z+H(z)/[k^2(1+z)\eta(z)]$
so that one obtains
\begin{equation}\label{eq:M_k_large}
  {\bf B}_{\bf k}(z)\simeq4\pi{\bf M}^\perp_{\bf k}(z)\,.
\end{equation}
This is the small-scale instantaneous limit in which the field
is given by the superposition of dipole fields which are only subject
to redshift since the second term in Eq.~(\ref{eq:M_k}) approximately
vanishes. In the opposite limit, $k\ll k_r$, the exponential
in Eq.~(\ref{eq:M_k_solution}) can be set to one and one obtains
\begin{eqnarray}\label{eq:M_k_small}
  {\bf B}_{\bf k}(z)&\simeq&4\pi{\bf k}^2(1+z)^2\int_z^\infty
  \frac{\eta(z^\prime){\bf M}^\perp_{\bf k}(z^\prime)}
  {(1+z^\prime)H(z^\prime)}dz^\prime\nonumber\\
  &\simeq&4\pi\left[\frac{k}{k_r(z)}\right]^2{\bf M}^\perp_{\bf k}(z)\,,
\end{eqnarray}
where in the second estimate we have used that
\begin{equation}\label{eq:M_evol}
{\bf M}^\perp_{\bf k}(z^\prime)\simeq\left(\frac{1+z^\prime}{1+z}\right)^2
{\bf M}^\perp_{\bf k}(z)
\end{equation}
In this limit the resulting field is thus frozen into the plasma and
strongly suppressed compared to the source term by a factor $(k/k_r)^2$ due
to the screening by the conductivity of the medium. 
It is easy to see that the critical wave number becomes smallest at late
times, when electrical conductivity and Hubble constant become smallest.
Using the Spitzer resistivity 
$\eta\simeq \pi m_e^{1/2}e^2/T^{3/2}$~\cite{choudhuri} one finds
\begin{equation}\label{eq:k_r}
  k_r(T)\simeq
  3\times10^{3}\left(\frac{T}{T_0}\right)^{1/2}
  \,{\rm pc}^{-1} \mbox{for $T\lesssim 1\,$eV}\, ,
\end{equation}
where $T_0$ is the present day CMBR temperature. This should be compared to,
for example, the comoving QCD Hubble radius wave vector, 
$k_{\rm QCD}\approx 6\,{\rm pc}^{-1}(T/100\, {\rm MeV})$, 
showing that screening of the plasma
is extremely efficient, leading only to highly suppressed magnetic fields
on large scales. The supression here, $n=7/2$ is substantially larger 
that that claimed to apply in Ref.~\cite{Caprini:2001nb}. 
However, we will see in the next section that
the conclusions change drastically when fluid motions are taken into account.

\section{Magnetic Field Evolution with Fluid Flows}

As post-inflationary "causal" magnetogenesis 
in the early Universe most likely takes place during
cosmic first-order phase transitions, and during such transitions the
generation of cosmic turbulence is inevitable, the complete
neglect of fluid motions is thus unrealistic. 
In the one-fluid approximation
of magnetohydrodynamics (MHD) the velocity field ${\bf v}$
evolves according to
\begin{equation}\label{eq:v}
  \partial_t{\bf v}+({\bf v}\cdot\nab){\bf v}=
  \frac{(\nab\times{\bf B})\times{\bf B}}{4\pi\rho}+{\bf f}_v\,,
\end{equation}
where $\rho$ is the fluid density and ${\bf f}_v$ is the viscous damping
force~\cite{Banerjee:2004df}.
We assume now that the fluid kinetic energy density (at least initially) dominates
magnetic energy density.
Furthermore we assume that the phase transition induced 
magnetization ${\bf M}$ has led
to the build-up of some magnetic seed field on relatively small scales
(cf. Sec. III), but
nothing on large scales. We are then interested to know how much magnetic
field will be generated on large scales due to non-linear interactions between
magnetic- and velocity- fields on smaller scales. 
The evolution of magnetic fields are given by 
\begin{equation}\label{eq:B1}
  \partial_t{\bf B}=\nab\times({\bf v}\times{\bf B})+\eta\Delta{\bf B}\, .
\end{equation}
We are not interested in the detailed magnetic field evolution, but rather
in the average buildup of magnetic energy density taken over an ensemble
of cosmic realizations, i.e.
\begin{equation}
\langle\partial_t M_k\rangle
\end{equation}
with $M_k$ the magnetic spectral energy (not to be confused
with the magnetisation) defined through 
\begin{equation}
\epsilon_B = \frac{1}{8\pi V}\int {\rm d^3}x\, {\bf B}^2(x) \equiv
\int {\rm d}k\, M_k 
\end{equation}
where $V$ denotes volume and where we have implicitly assumed cosmic
homogeneity and isotropy rendering $M_k$ a function only of the magnitude
of wavevector $\bf k$.

Ensemble averaging over the product of Eq.~(\ref{eq:B1}) and $\bf B$,
assuming time-independent and uncorrelated fluid turbulence, i.e.
$\langle v^*_{\rm k}(t)v_{\rm k^{\prime}}(t^{\prime})\rangle
\sim \delta_{\bf k^{\prime}k}\delta_{t^{\prime}t}$ unmodified by the
magnetic fields, i.e. neglecting the backreaction magnetic force term
in Eq.~(\ref{eq:v}. Such a calculation has been performed first by 
Ref.~\cite{Vainshtein}. We follow here the analysis of 
Ref.~\cite{Kulsrud} which leads to the
following evolution equation for $M_k$~\cite{remark}
\begin{equation}
\label{eq:master0}
\frac{\partial M_q}{\partial t}\simeq {\cal C}\int {\rm d}k{\rm d \theta}\,
\frac{q^4}{k\,k_1^3}\,\frac{v^2(L_{k_1})}{v(L_k)}f(q/k,\theta )M_k\, ,
\end{equation}
where for simplicity we have dropped the ensemble average brackets
and discarded terms of higher order in wavevector $\bf k$ as well as
terms containing magnetic helicity. In other words, we are only interested
in the small $k$ tail of non-helical fields. Eq.~(\ref{eq:master0}) describes
the generation of magnetic energy on scale with wave vector  
$\bf q$ by the non-linear interaction between magnetic modes with
wave vector $\bf k$ and velocity modes with wave vectors 
${\bf k_1}\equiv {\bf q} - {\bf k}$. Here $\theta$ is the angle between
$\bf q$ and $\bf k$, $v(L_k)$ are typical velocities on scale $L_k = 2\pi/k$ 
associated with wave vector $k$, and $\cal C$ is a crucial constant 
estimated approximately as ${\cal C}\sim (2\pi)^2$.  Furthermore, the
function $f$ is given by 
\begin{equation}
f(x,\theta ) = {\rm sin^3}(\theta )\, 
\frac{(1+x^2-x{\rm cos}\theta)}{(1+x^2-2x{\rm cos}\theta)}\, , 
\end{equation}
where the denominator is $k_1/k$.
We now assume a power law for the scale dependence of the fluid velocity, i.e.
\begin{equation}
\label{eq:valpha}
v^2(L_k) = v_{g}^2\biggl(\frac{L_0}{L_k}\biggr)^{\alpha} =
v_{g}^2\biggl(\frac{k}{k_0}\biggr)^{\alpha}\, ,
\end{equation}
where $L_0 = 2\pi/k_0$ is some reference scale on which $v = v_{g}$.
Here $\alpha = 3$ corresponds to uncorrelated white noise whereas 
$\alpha = -2/3$ corresponds to a red Kolmogoroff spectrum.
Inserting Eq.~(\ref{eq:valpha}) into Eq.~(\ref{eq:master0}), changing to
comoving wave vectors $k$ with $k_{\rm phys}=k/a$, where $a$ denotes the scale
factor, one finds
\begin{equation}
\label{eq:master1}
\frac{\partial M_q}{\partial {\rm\, ln}a}\simeq {\cal C} \frac{a}{H_g} 
\int_q^{k_I(a)} {\rm d\, ln}k\, k\biggl(\frac{q}{k}\biggr)^4v_{g}
\biggl(\frac{k}{k_0}\biggr)^{\alpha/2}G^{\alpha}\biggl(\frac{q}{k}\biggr)\, M_k
\end{equation}
for the evolution of magnetic energy density during a radiation dominated
Universe. Here $k_I(a)$ is some effective ultraviolet cutoff, or damping 
scale, to be discussed below, and $H_g$ is the Hubble constant at the initial
magnetogenesis era where we have chosen scale factor $a=1$. Note that
equation Eq.~(\ref{eq:master1}) really holds for some properly defined comoving
$M_k$ which takes into account the redshifting $\epsilon_B\sim 1/a^4$
of magnetic energy density~\cite{remark2}. During matter domination
the dependence on the scale factor of Eq.~(\ref{eq:master1})
is changed from $a$ to $a^{1/2}$. The function $G^{\alpha}$ in 
Eq.~(\ref{eq:master1}) contains the $\theta$ integration
\begin{equation}
\label{eq:G}
G^{\alpha}(x) = \int_0^{\pi}{\rm d}\theta\, {\rm sin^3}(\theta )
\frac{(1+x^2-x{\rm cos}\theta)}{(1+x^2-2x{\rm cos}\theta)^{5/2-\alpha/2}}\, .
\end{equation}
We are only interested in large-scale modes $q\lsim k$. In this case 
$G^{\alpha}(x)$ is a slowly varying function, with, for example $G^3$
varying only from $4/3$ to $5/3$ when $x$ goes from 0 to 1. We therefore 
conservatively drop it. 

Eq.~(\ref{eq:master0}) represents our master equation. It demonstrates that
magnetic energy may be transferred from small scales (large $k$) to large
scales (small $q$) via non-linear processes between fluid flows and magnetic
fields, and that this process apparently becomes more efficient at later epochs (larger $a$). 
Before presenting a prediction for the evolution of the 
large-scale tail of magnetic fields, we have to emphasize a conceptual
difference between the findings of this work and all other prior studies
of primordial magnetic fields. So far, it had been believed that the epochs of
magnetogenesis and subsequent field decay are well separated. In other
words, it
would be assumed that some magnetogenesis scenario during a cosmic phase 
transition would fully account for all field properties, in particular, also
those on large scales, and that the final late time field would be simply
given by all magnetic energy which hadn't been dissipated yet. This study shows that
magnetogenesis {\it should} be considered as a continous process, which in
particular holds true for the large-scale tail, 
with magnetic fields generated on small $q$ even long after the epoch of 
the initial magnetogenesis era.

We now proceed to estimate the growth of large-scale magnetic fields. 
We assume that the kinetic energy density is peaked on some scale $k_I$, 
which we refer to as the integral scale. For $k > k_I$ the velocity 
spectrum is red (e.g. Kolmogoroff), and for $k < k_I$ given by 
Eq.~(\ref{eq:valpha}). Contributions from $v$ modes with $k> k_I$
to the integral Eq.~(\ref{eq:master0})
may be approximatively neglected, as already done in Eq.~(\ref{eq:master1}).
We assume that the magnetic seed field 
only exists on some small scale $k_B$, with $M_q = 0$ for either
$q\ll k_B$ and $q\gg k_B$. For the moment we make the crucial assumption that
$k_B\sim k_I$ and that equipartition holds on that scale, 
i.e. $M_k \approx E_k$, with $\epsilon_K \equiv \int {\rm d\,}k E_k$, where
$\epsilon_K$ is the kinetic energy density. We will discuss this assumption in
some more detail below.

We need a model of the evolution of fluid flows in the expanding Universe,
as this modifies the $v(L_k)$ entering Eq.~(\ref{eq:master1}).
Such a model has been presented, in detail, in Ref.~\cite{Banerjee:2004df}. 
It had been found
that the evolution of velocity and magnetic fields in the early Universe
is quite complex, described by an alternation of turbulent phases and viscous
phases. Here, early on, viscosity is due to neutrinos, and at later stages,
due to photons. Nevertheless, we focus first on turbulent phases.

\subsection{Turbulent Phases}

In the absence of further turbulence producing mechanisms, 
turbulent flows simply decay when the eddy turnover rate equals the
Hubble rate, i.e.
\begin{equation}
\label{eq:integral}
\frac{v^p(k_I)}{L^p_I} = \frac{v^p(k_I)\,k^p_I}{2\pi}\simeq H\, . 
\end{equation}
In particular, eddies on $k_I$ break up into eddies of larger $k$, with 
their energy getting transported to the dissipative scale 
$k_{\rm diss}>k_I$ where it is converted to heat. Here the superscript $p$
denotes proper quantities. These are related to comoving quantities via
$L^p=L\, a$, $k^p=k/a$, as well as $v^p = v$ (radiation dominated) and
$v^p = v/a$ (matter dominated). From Eq.~(\ref{eq:integral}), using
Eq.~(\ref{eq:valpha}) and assuming radiation domination one derives
\begin{equation}
\label{eq:kintegral}
k_I(a) = k_I^0\,a^{-2/(\alpha+2)}\, .
\end{equation}
with
\begin{equation}
k_I^0 = \frac{2\pi H_g}{v_g}
\end{equation}
the initial integral scale.
Here we adopted the reference scale $k_0 = k_I^0$ such that
$\epsilon_K^g = \rho_r^gv_g^2/2$ represents the initial turbulent energy
density.  

Assuming equipartition $M_k(k_I^0)\approx E_k(k_I^0)$ 
on the initial integral scale $k_I^0$ but $M_k\approx 0$ beyond,
we are interested to know if the decaying magnetohydrodynamic turbulence 
may establish equipartition on even larger scales, and if yes, how large 
is the largest scale where equipartion may be achieved. Using 
Eq.~(\ref{eq:kintegral}), assuming equipartition $M_k\approx E_k$
on $k_I$ at all times (to be verified selfconsistently below), 
one may derive for the change in magnetic energy density $\Delta M_q(a)$ at
redshift $a$
\begin{equation}
\label{eq:DeltaM}
\Delta M_q(a) \simeq {\cal C}\rho_r
\biggl(\frac{q}{2\pi H_0}\biggr)^4
\frac{v_{g}^7}{H_0}
a^{\frac{10-2\alpha}{\alpha +2}}\, ,
\end{equation}
where $\rho_r$ is radiation energy density and where
we neglected an unimportant numerical factor. The total change
of $M_q$ is conservatively determined at $a_q$ where the integral scale 
has decreased to $q$, i.e. $k_I(a_q) = q$. The condition that by 
$a_q$ the change 
$\Delta M_q(a_q) > M_q^{\rm equi} = v_{g}^2(q/k_I^0)^{\alpha}/q$ is larger
than the equipartition value then yields
\begin{equation}
{\cal C}\gsim \frac{1}{2\pi}
\end{equation}
independent (!) of $q$, $v_g$, $H_g$, and $\alpha$. As we determined
${\cal C}\approx (2\pi )^2$ thus equipartition between magnetic
fields and velocity fields may be attained on a large range of scales, 
independent of the turbulence spectrum. This justifies selfconsistently 
our above assumption.

\begin{figure}
\epsfxsize=8.5cm
\includegraphics[width=0.45\textwidth]{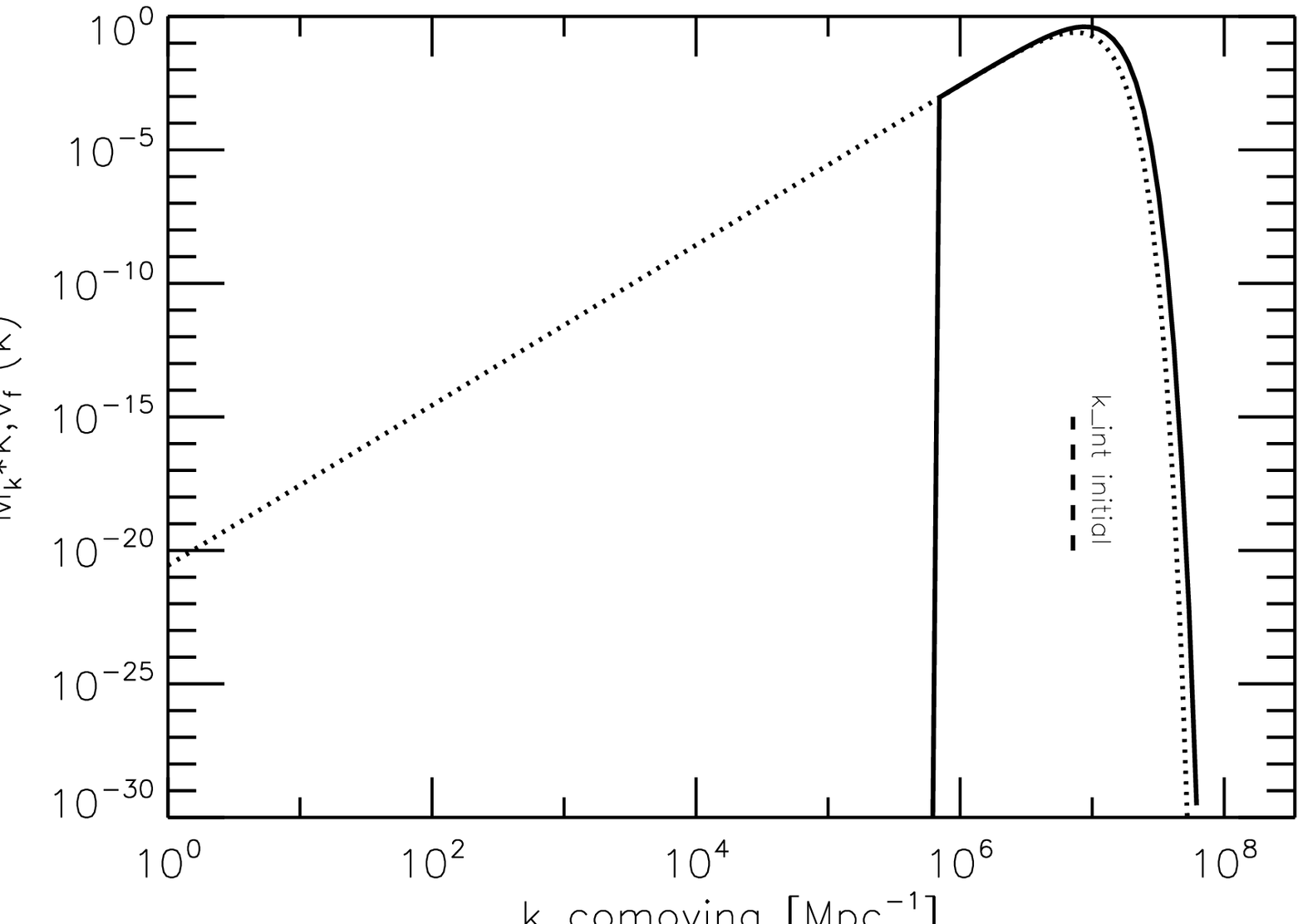}
\includegraphics[width=0.45\textwidth]{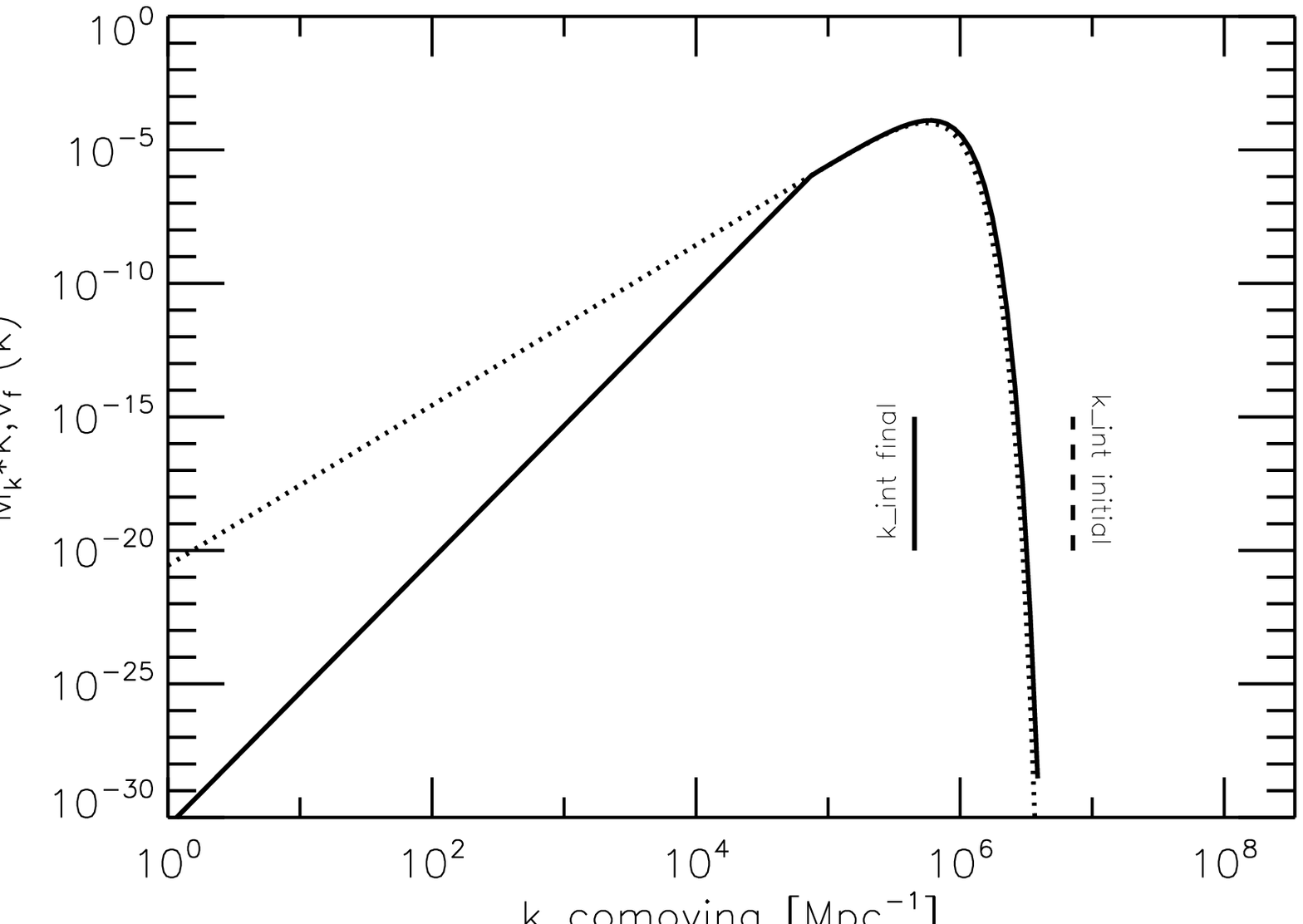}
\includegraphics[width=0.45\textwidth]{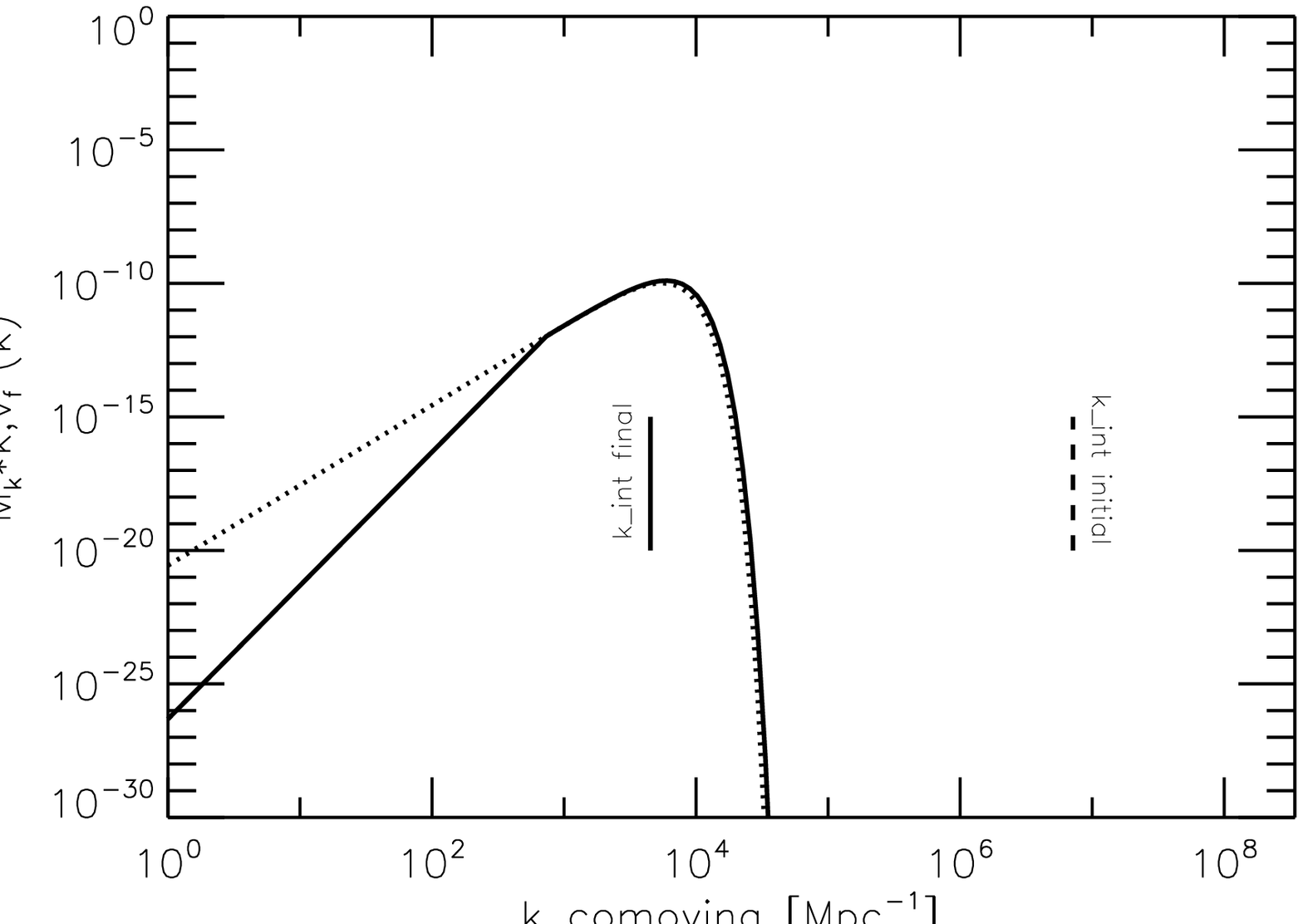}
\caption{
Evolution of the magnetic energy spectrum (in units of
the radiation energy density) according to
Eq.~(\ref{eq:master1}) for a magnetogenesis temperature $T=100\,$MeV
and for $v_g=1$ and $\alpha=3$. The upper panel shows the assumed 
initial condition for the magnetic- (solid) and kinetic- (dotted) 
energy spectrum at $T=100\,$MeV, whereas the middle and lower panels
show these quantities evolved to $T=100\,$keV and $T=1\,$eV, respectively. 
Dashed and solid vertical line
represent the integral scale at the initial and current temperature, 
respectively. The scale at
which the solid line departs from the dotted line, is the scale
$k_{\rm equi}$ from Eq.~(\ref{eq:keq}) which in the present case reads
$k_{\rm equi}(a)\simeq0.063k_I(a)$. This is nicely confirmed by the
simulation. Note that the semianalytic simulation assumes a turbulent
regime throughout.} 
\label{fig1}
\end{figure}

At any late time, the magnetic field spectrum therefore has the form
\begin{equation}\label{eq:spectrum}
  \frac{M_q(a)}{M_I(a)}=\left\{\begin{array}{lll}
  \left[\frac{q}{k_I(a)}\right]^{-\beta} 
& \mbox{$k_I(a)\leq q \leq k_{\rm diss}(a)$}\\
  \left[\frac{q}{k_I(a)}\right]^{\alpha -1}
& \mbox{$k_{\rm equi}(a)\leq q \leq k_{I}(a)$}\\
  \left[\frac{k_{\rm equi}(a)}{k_I(a)}\right]^{\alpha -1}
       \left[\frac{q}{k_{\rm equi}(a)}\right]^{4}
& \mbox{$q\leq k_{\rm equi}(a)$}\,,\end{array}\right.
\end{equation}
with the $q-$independent constant $M_I(a)=M^{\rm equi}_{k_I(a)}$. This is
Kolmogoroff ($\beta=2/3$) or Ironshnikov-Kraichnan ($\beta=1/2$) 
on very small scales, following the fluid spectrum on scales larger 
than the integral scale $k_I(a)$ and turning over to a steeper spectrum 
for $k<k_{\rm equi}(a)$. Here $k_{\rm equi}(a)$ is the smallest wavevector for which
equipartition could be attained at redshift $a$. It is given by
\begin{equation}
\label{eq:keq}
k_{\rm equi}(a) \simeq (2\pi {\cal C})^{\frac{1}{\alpha -5}}\, k_I(a)\,.
\end{equation} 
The spectral index $4$ for $k<k_{\rm equi}$ in Eq.~(\ref{eq:spectrum})
is dictated by the power of $q$ in Eq.~(\ref{eq:master1}).
The evolution of the spectrum for a representive case is
displayed in Fig.~\ref{fig1}, with the spectrum
shown at temperatures $T=100\,$MeV, $T=100\,$keV, and $T=1\,$eV, 
respectively. Note that since the integral in the master
equation~(\ref{eq:master1}) is dominated
by the integral scale, these results are insensitive to the turbulence
power spectrum at scales $k\ll k_I(a)$, and thus in particular on
whether this spectrum is causal or not.

Eq.~(\ref{eq:kintegral}) and Eq.~(\ref{eq:DeltaM}) apply only during
radiation domination. It is easy to show~\cite{Banerjee:2004df} 
that during matter 
domination $k_I(a)$ stays constant or at best decreases logarithmically 
with scale factor. Using this plus the matter-domination modified
Eq.~(\ref{eq:master1}), as well as taking into account of the proper
redshifting
of $v$, it can be shown that no further substantial growth of
$M_q$ occurs during matter-domination. This holds true even when 
$v\approx v_A$, as should be the case, yielding a milder redshift 
dependence $v\sim a^{-1/2}$ than the above stated $v\sim a^{-1}$.
Here $v_A$ is the Alfven velocity. 
This implies that $k_{\rm equi}$ gets "frozen in" at the transition of
radiation to matter domination at its minimum value $k_{\rm equi}^{\rm min}$. 

We comment here shortly on one implicit assumption made. In particular,
we allow $M_q$ only to grow to its scale-dependent equipartition value
$M_q\approx E_q$. Though this seem fairly plausible, it is not clear if,
in fact, it is true. Rather, it is possible that on scales $q < k_I$
the magnetic field grows to super-equipartition, i.e. $M_q > E_q$. 
This may be possible as long as global energy conservation is not violated.
It is
not clear if this could happen, and only an improved analysis taking 
account of potential modifications of Eq.~(\ref{eq:master0}) due to
backreaction of magnetic fields on fluid flows, as well as deriving an
independent equation for $\langle \partial_t E_k\rangle$, could 
conclusively adress this issue. This formidable task is however beyond
the scope of this paper. We note that our assumption $M_q\lsim E_q$ 
is conservative.

\subsection{Viscous Phases}

A realistic calculation has to take into account the various small
Reynolds number $R<1$ viscous phases of magnetohydrodynamics in the
early Universe. Here viscous forces in Eq.~(\ref{eq:v}) are given by
~\cite{Banerjee:2004df}
\begin{equation}
{\bf f}_v = \left\{\begin{array}{ll}
\eta {\bf\nabla}^2{\bf v} & l_{\rm mfp}\ll L\\
-\alpha {\bf v} & l_{\rm mfp}\gg L
\end{array}\right.\,,
\end{equation}
where $l_{\rm mfp}$ is the mean free path of the momentum transporting
particles, early on neutrinos, and later photons, and $L$ is the scale
under consideration. It can be shown that the diffusion constant $\eta$ 
and drag constant $\alpha$ are essentially $\eta\simeq l_{\rm mfp}$ and
$\alpha \simeq 1/l_{\rm mfp}$ up to numerical, and statistical weight 
factors. Starting at high temperatures, as the mean free path of 
neutrinos increases with the
expansion of the Universe, an initially turbulent fluid becomes 
viscous due to neutrino diffusion. At this point all pre-existing
fluid motions will be erased up to the neutrino  
diffusion length, i.e. $d\simeq \sqrt{\eta t}$ with $t$ cosmic time. 
However, importantly, magnetic fields are not erased, and remain frozen 
into the plasma. On the other hand build-up of large-scale magnetic fields 
via Eq.~(\ref{eq:master0}) has stopped entirely as there are no appreciable 
fluid velocities. This viscous phase occurs already at fairly high 
temperatures $T\sim 20-80\,$MeV, depending on the kinetic- and 
magnetic- energies. As the neutrino mean free path increases beyond the
scale $L_I\sim 1/k_I$, the fluid continues its viscous period in the neutrino 
free-streaming regime. Form their on, the very small Reynolds number starts 
increasing again. Here secondary fluid velocities 
\begin{equation}
\label{eq:v_visc}
v\simeq \frac{k_I\,B^2}{\rho\alpha} \simeq \frac{v_A^2}{\alpha L_c}\, ,
\end{equation}
are generated via the magnetic stresses in Eq.~(\ref{eq:v}). These
velocities become larger and larger, and at some point, still in the viscous
regime, the "instantaneous" dissipation scale $k_{\rm diss}$
falls below the integral scale $k_I$ at the beginning of the 
viscous regime. At this point further magnetic- and fluid- energy density
is dissipated leading to an increase of $k_I\approx k_{\rm diss}$.
Here $k_{\rm diss}$ is given again by the condition 
Eq.~(\ref{eq:integral}) with velocities now as in Eq.~(\ref{eq:v_visc}). 
Somewhat later, typically at $T\sim 5-10\,$MeV shortly before neutrino
decoupling, the fluid becomes again fully turbulent. This cycle repeats 
itself at temperatures $T\lsim 1\,$MeV, but now due to the long mean free 
path of photons.

\begin{figure}
\epsfxsize=8.5cm
\includegraphics[width=0.5\textwidth]{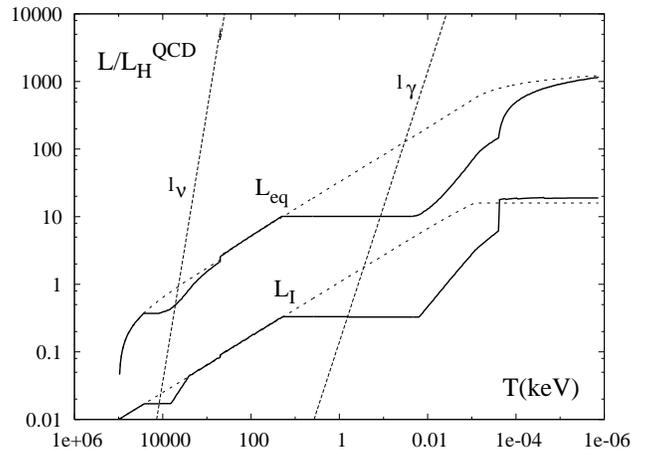}
\caption{Evolution of various length scales as a function of temperature.
All length scales are in units of the approximate QCD Hubble radius, i.e.
$L_H^{QCD}\approx 1\,$pc. Shown are the integral scale $L_I$ on
which most of the magnetic- and kinetic- energy is contained, as well as
the equipartition scale $L_{\rm equi}$, i.e. the largest scale on which equipartition
between kinetic turbulent energy and magnetic energy could be attained. Each
length scale is shown twice, once for a Universe which is always turbulent
(dotted lines) and once for a realistic Universe, where turbulent and
viscous MHD are alternating due to neutrino- and photon 
viscosity (solid lines). The dashed lines, show the neutrino- and photon- mean
free path as labeled. The results of this semi-analytic simulation assume
an initial magnetogenesis epoch at $T=100\,$MeV with 
$\epsilon_K^g = 10^{-4}\rho_r^g$ and $\epsilon_B^g = 10^{-9}\rho_r^g$,
where $\epsilon_B^g$, $\epsilon_K^g$, and $\rho_r^g$ are initial magnetic-,
kinetic-, and total radiation energy density, respectively. A 
white noise $\alpha =3$ spectrum has been assumed for the slope of the
turbulence on large scales $L\gsim L_I$.}
\label{fig2}
\end{figure}
 
It is known~\cite{Banerjee:2004df} 
that the net effect of theses viscous phases is a
delay of the growth of the integral scale $L_I(a)$ during viscous periods
themselves. Nevertheless, once turbulence has recommenced, the integral
scale $L_I(a)$ is indistinguishable from $L_I^{\rm turb}(a)$, 
the respective scale in a Universe which had always remained turbulent.
This may be observed in Fig.~\ref{fig2}, where results for $k_I(a)$ of a 
semi-analytic calculation of $L_I$ are shown. The simulation takes proper 
account of the correct neutrino- and photon- viscosities.
It is far from clear if primordial viscosity has a similar effect on 
$L_{\rm equi}(a)$ the maximum scale where local equipartition on large scales 
has been attained, in particular, simply the delay of growth of this 
quantity. During large parts of the viscous eras no growth of magnetic
field energy density via Eq.~(\ref{eq:master0}) may occur. This could yield
much weaker $M_q$. On the other hand, towards the end of the viscous periods
$M_{k_I}$ entering Eq.~(\ref{eq:master0}) is genuinely larger than when
viscosity would be small. It turns out that both effects roughly cancel each
other as can be seen in 
Fig.~\ref{fig2}. The detailed simulation shows that as
in the case of $k_I(a)$, $k_{\rm equi}(a)$ during turbulent phases
is as if viscous phases had never occurred.

We now briefly comment on the adopted initial conditions of equipartition
on $k_I$, though we do not intend to go into much detail of the initial
magnetogenesis period itself. Using Eq.~(\ref{eq:master0}) one may derive
an approximate equation $\partial M_{k_I}/\partial t\simeq M_{k_I}/\tau$
with $\tau$ given by $\tau^{-1}\approx {\cal C}v(k_I)k_I$ and thus
is a fraction of the eddy turnover time on the integral scale. We note that since
$v(k_I)k_I/(2\pi)\gsim H$ the requirement to have more than one e-fold
of amplification of magnetic fields on $k_I$ per Hubble time during the
phase transition reads ${\cal C}\gsim 2\pi$. Since this is the case, even
initially much weaker than equipartion fields are likely to come 
to equipartion during the phase transition. Nevertheless, this does not
adress concerns about possibly most magnetic energy going into large $q$
modes, due to the stronq $q^4$ dependence in Eq.~(\ref{eq:master0}).

\subsection{Results} 

Given all the above, the final
present-day spectrum, barring further evolution in the non-linear structure
formation regime, is therefore given by Eq.~(\ref{eq:spectrum}) with
$k_{\rm equi}$ and $k_I$ replaced by $k_{\rm equi}^{\rm min}$ and $k_I^{\rm min}$, their
respective values at matter-radiation equality. In this subsection we 
deviate from our above chosen convention that $a=1$ at the initial
magnetogenesis epoch. Rather, as customary, we choose $a=1$ at the present 
epoch. In this case $k_I^{\rm min}$ and $M_I(a_{eq})$ in 
Eq.~(\ref{eq:spectrum}) are given by
\begin{eqnarray}
\label{eq:kImin}
k_I^{\rm min} = k_I(a_{eq})  & = & 
\frac{2\pi H_{g}}{v_g}
\biggl(\frac{H_{eq}}{H_g}\biggr)^{\frac{\alpha+4}{2\alpha+4}}
a_{eq}\\
\label{eq:MqI}
M_I(a_{eq}) & = & {\rho_rv_g^2}\,
\biggl(\frac{H_{eq}}{H_g}\biggr)^{\frac{2\alpha}{\alpha+2}}
\frac{1}{k_I(a_{eq})}\, ,
\end{eqnarray}
where $H_{eq}$ and 
$a_{eq}\approx 5\times 10^3$ are Hubble constant
and scale factor at matter radiation equality, respectively. 
We note that for $v_g\simeq 1$, i.e. the speed of light, their is
equipartition between kinetic turbulent energy $\epsilon^g_K$ 
and radiation density $\rho^g_r$ (as well as
subsequently magnetic energy density $\epsilon^g_B$). Such
equipartion corresponds to
$B_0\simeq 5.7\times 10^{-6}\,$Gauss in comoving magnetic field strength.
Making the 
reasonable assumption that the turbulence is white 
noise (cf.~\cite{Kahniashvili:2010gp} for a numerical justification), i.e. 
$\alpha = 3$ yields spectral index $n=3/2$ for $k>k_{\rm equi}$ and only for
$k < k_{\rm equi}$ the in Ref.~\cite{Caprini:2001nb} proposed $n=5/2$. We will see  
that this leads to strongly modified predictions for the possible
large-scale magnetic field strength arising due to "causal" processes
compared to those given in Ref.~\cite{Caprini:2001nb}.
The scale $L_{\rm equi}^{\rm min}=2\pi/k_{\rm equi}^{\rm min}$ may be obtained via 
Eq.~(\ref{eq:keq}) and Eq.~(\ref{eq:kImin})  
\begin{equation}
\label{eq:Leqmin}
L_{\rm equi}^{\rm min} \simeq 30\,{\rm kpc}\, 
\biggl(\frac{\epsilon^g_K}{\rho^g_r}\biggr)^{1/2}
\biggl(\frac{T_g}{100 {\rm MeV}}\biggr)^{-3/5}\, 
\quad \mbox{for $\alpha = 3$}\, .
\end{equation}
To derive an upper bound we assume 
the maximum plausible energy in the turbulence, i.e. equipartition 
$\epsilon^g_K\simeq \rho^g_{r}$ with the radiation energy density. 
We derive the magnetic field strength on scale $L$ from 
Eq.~(\ref{eq:spectrum}) and Eq.~(\ref{eq:kImin}) - Eq.~(\ref{eq:MqI})
taking account of the conversion given below Eq.~(\ref{eq:MqI}).
On $L\simeq 100\,$kpc, the scale chosen by Ref.~\cite{Caprini:2001nb} we find
$B(100 {\rm kpc})\lsim 5\times 10^{-14}\,$Gauss for the QCD transition 
($T_g\simeq 100\,$MeV). This value is also reflected by
Fig.~\ref{fig1}, lower panel when using that
$B_k\simeq5.7\times10^{-6}\,(M_kk/\rho_r)^{1/2}\,$Gauss.  
For the electroweak transition ($T_g\simeq 100\,$GeV) one has $B(100
{\rm kpc})\lsim 3\times 10^{-20}\,$Gauss. This should be compared
to the much smaller $B(100 {\rm kpc})\lsim 10^{-27}\,$Gauss given in 
Ref.~\cite{Caprini:2001nb} for the electroweak transition. 

However, we stress here 
again~\cite{Banerjee:2003xk,Banerjee:2004df} that more important than 
the magnetic field strength on some large scale $\sim 100\,$kpc should be
the total magnetic energy on all scales. Numerical simulations 
show~\cite{Dolag:2002bw} that gravitationally induced inverse
cascades during cluster collapse seems to transfer this magnetic power 
to large scales. We thus restate here~\cite{Banerjee:2003xk,Banerjee:2004df} 
the prediction for present day magnetic field strength and coherence length 
as a function of magnetogenesis temperature $T_g$, initial 
magnetic/turbulent energy density, $r_g$
(in units of radiation energy density), and initial helicity $h_g$ (as a
fraction of maximal helicity)
\begin{eqnarray}
L_c (T) & \simeq & 12\, {\rm pc} \,\,
   \left(\frac{r_g}{0.01}\right)^{1/2} \,
   \left(\frac{T_g}{100\,\rm GeV}\right)^{-3/5} 
\label{eq:Lc0exnohel}
\nonumber
\\ 
B_c(T) & \simeq & 6.0\times 10^{-14} \, {\rm G} \, \,
    \left(\frac{r_g}{0.01}\right)^{1/2} \,
   \left(\frac{T_g}{100\,{\rm GeV}}\right)^{-3/5}
\label{eq:Bc0exnohel}
\end{eqnarray}
for non-helical fields and
\begin{eqnarray}
L_c (T) & \simeq & 1.9\, {\rm kpc}\,\, 
   \left(\frac{r_g}{0.01}\right)^{1/2} \,
   \left(\frac{h_g}{0.01}\right)^{1/3} \,
   \left(\frac{T_g}{100\,\rm GeV}\right)^{-1/3}\, 
\label{eq:Lc0exhel}
\nonumber
\\
B_c(T) & \simeq & 1.6\times 10^{-11}\,{\rm G}\, \, \\ & &
    \times\left(\frac{r_g}{0.01}\right)^{1/2} \,
   \left(\frac{h_g}{0.01}\right)^{1/3} \,
   \left(\frac{T_g}{100\,\rm GeV}\right)^{-1/3}\, ,
   \nonumber
\label{eq:Bc0exhel}
\end{eqnarray}
for helical fields. Here Eq.~(\ref{eq:Bc0exnohel}) assumes a 
spectral index $n=3/2$.

\section{Conclusions}

In this paper we have considered the large-scale (small wavenumber $k$) 
tail of primordial magnetic fields produced "causally" (i.e. not during
an inflationary epoch) in the early Universe. Here we focussed on scales
which are instantaneously super-Hubble during the initial magnetogenesis
epoch, i.e. $\lambda\gsim 1\,$pc, as such scales correspond to current 
galactic scales. It has been claimed~\cite{Caprini:2001nb} 
that due to causality such fields are
highly supressed. i.e. $n=5/2$ in Eq.~(\ref{eq:n}). We have shown here that
this may not be the case.

It is customarily assumed that an initial period of 
magnetogenesis (e.g. during a phase transition) may be well separated from
the subsequent epoch of freely decaying MHD. In other words, field production
only happens during the magnetogenesis epoch, whereas subsequently only field
dissipation takes place. However, this is not the case, magnetic field on
large scales may be continously generated due to nonlinear interactions
of small scale fluid- and magnetic- modes, even well after the initial epoch
of magnetogenesis. We have shown that, with the expansion of the Universe, 
this process is rapid enough to yield scale-dependent equipartition between 
fluid velocities and magnetic fields on ever and ever larger scales.
In particular, barring super-equipartion values of the magnetic field, this
implies that the kinetic- and magnetic- energy tail could have the same spectral
index. As white noise $n=3/2$ for the fluid is 
expected~\cite{Kahniashvili:2010gp}, the magnetic field
tail could also develop $n=3/2$. 

When the realistic alternation of turbulent and viscous MHD 
(due to photon- and neutrino- fluid viscosity)
in the early Universe is considered, the evolution becomes 
substantially more complicated. We have followed
this with a semi-analytic treatment. During turbulent epochs
the fluid forces equipartition with magnetic fields on small scales. 
After this, amplification of large-scale magnetic fields due to non-linear
interactions between velocity- and magnetic- modes on small
scales becomes important. When
fluid viscosity becomes considerable, all fluid flows up to the 
neutrino (photon) diffusion scale become erased, and the process of 
amplification of large-scale magnetic fields effectively stops. Only when 
neutrino (photon) interaction with the plasma become very weak, new fluid 
flows are generated by the priorly generated and frozen in magnetic fields.
Thus the fluid initial kinetic energy had been stored in magnetic field energy. 
Subsequently further large-scale magnetic field amplification takes place.
As complicated as is, the net result of these processes is a present day
large-scale magnetic field tail essentially indistinguishable from that
if the Universe would have stayed turbulent throughout.

However, we caution that our results are
preliminary, as our evolution equation may not include all magnetic sink terms
on large scales. Those were assumed to become important only
at scale-dependent equipartition, i.e. when $M_q\approx E_q$. Furthermore,
backreaction of magnetic fields on fluid flows was only taken into account in
a heuristic way. A complete analysis, which is beyond the scope of the
present paper, should be performed to confirm our results. 

The resulting magnetic field spectrum may thus be 
$n=3/2$ up to some large scale 
$L_{\rm equi}^{\rm min}$ in Eq.~(\ref{eq:Leqmin}) and
$n=5/2$ beyond. This would lead to considerably larger magnetic field strength
on large scales than claimed in Ref.~\cite{Caprini:2001nb}.
If appreciable primordial magnetic fields actually exist, however, 
can be verified only  by measuring the magnetic field in the 
intergalactic medium.

\vskip 0.15in
{\it Acknowledgments}
We thank Chiara Caprini, Torsten Ensslin, Ruth Durrer, 
Tina Kahniashvili, Kandu Subramanian, and Tanmay Vachaspati 
for useful conservations.
This work was supported by the Deutsche Forschungsgemeinschaft through SFB 676
``Particles, Strings and the Early Universe: The Structure of Matter and
Space-Time''.


\begin{thebibliography}{99}
\bibitem{remark}
The equation follows from Ref.~{}'s Eqs. (2.30) (with only the first term on 
the RHS), Eqs. (2.31), (2.33), and (2.34). The conversion of $J(k,0)$ in their
Eq. (2.34) to fluid velocity may be obtained from Eqs. (3.1) and (3.4).
\bibitem{remark2}
One may use, for example, $\tilde{M}_k \equiv M_k/s^{4/3}$, provided
the integral $\int {\rm d}\,k\, M_k$ is performed over comoving wave vectors.
Here $s$ is radiation entropy with $s\, a^3 =\,{\rm const}$.
\end{thebibliography}
\end{document}